# DeepCOVID-Fuse: A Multi-modality Deep Learning Model Fusing Chest X-Radiographs and Clinical Variables to Predict COVID-19 Risk Levels

Yunan Wu[1], Amil Dravid[1], Ramsey Michael Wehbe[2], Aggelos K. Katsaggelos[1]

[1] Department of Electrical and Computer Engineering, Northwestern University, Evanston, IL, 60201, USA

[2] The Division of Cardiology, Department of Medicine and Bluhm Cardiovascular Institute, Northwestern Memorial Hospital, Chicago, 60611, USA

corresponding to **Yunan Wu** (yunanwu2020@u.northwestern.edu)

# Abstract

**Propose:**
To present DeepCOVID-Fuse, a deep learning fusion model to predict risk levels in patients with confirmed coronavirus disease 2019 (COVID-19) and to evaluate the performance of pre-trained fusion models on full or partial combination of chest x-ray (CXRs) or chest radiograph and clinical variables.

**Materials and Methods:**
The initial CXRs, clinical variables and outcomes (i.e., mortality, intubation, hospital length of stay, ICU admission) were collected from February 2020 to April 2020 with reverse-transcription polymerase chain reaction (RT-PCR) test results as the reference standard. The risk level was determined by the outcome. The fusion model was trained on 1657 patients (Age: 58.30 ± 17.74; Female: 807) and validated on 428 patients (56.41 ± 17.03; 190) from Northwestern Memorial HealthCare system and was tested on 439 patients (56.51 ± 17.78; 205) from a single holdout hospital. Performance of pre-trained fusion models on full or partial modalities were compared on the test set using the DeLong test for the area under the receiver operating characteristic curve (AUC) and the McNemar test for accuracy, precision, recall and F1.

**Results:**
The accuracy of DeepCOVID-Fuse trained on CXRs and clinical variables is 0.658, with an AUC of 0.842, which significantly outperformed ($p < 0.05$) models trained only on CXRs with an accuracy of 0.621 and AUC of 0.807 and only on clinical variables with an accuracy of 0.440 and AUC of 0.502. The pre-trained fusion model with only CXRs as input increases accuracy to 0.632 and AUC to 0.813 and with only clinical variables as input increases accuracy to 0.539 and AUC to 0.733.

**Conclusion:**
The fusion model learns better feature representations across different modalities during training and achieves good outcome predictions even when only some of the modalities are used in testing.

## Introduction

Coronavirus disease 2019 (COVID-19) has been heavily straining the healthcare systems of countries across the world, with over 500 million cases and 6 million deaths as of July 2022 (1). Reverse-transcription polymerase chain reaction (RT-PCR) is the current standard for the diagnosis of COVID-19. It is typically used by authorized, trained clinical laboratory personnel and patients with suspected COVID-19. Additionally, results can take over twenty-four hours (2). However, previous studies have demonstrated the detection of COVID-19 features in chest radiographs (X-rays) and CT images (3), which can be combined with clinical judgment to make COVID-19 diagnosis. Artificial Intelligence (AI) shows promise in detecting these signatures with chest X-rays and CT scans (4–7), enabling faster and more accurate treatment of suspected patients. The use of CT in this domain is restrictive, particularly due to cost and time, whereas X-ray scanning is more commonly conducted and accessible. Nevertheless, the use of these scans is not discriminative enough to serve as a diagnostic tool, but can instead be used to inform the diagnosis and flag at-risk patients who may need further testing (8). Especially in places with limited resources, the use of chest X-rays can aid in resource allocation, triage, and infection control.

Many AI models have been proposed for the purpose of COVID-19 detection (9–12). However, their use of small and biased public datasets drives the skepticism for their deployment (13). DeepCOVID-XR, which proposes an ensemble of convolutional neural networks (CNNs), serves as one of the first works trained and tested on a large clinical data set for COVID-19 detection (4). It has been found to be more robust to biases present in models trained on public data. With rapid testing more readily available than it was at the beginning of the pandemic, more efforts can be made to serve those who are infected. Identifying the severity and prognosis of infected individuals can further aid in triaging and resource allocation. In fact, most published AI algorithms focused on better detection of COVID-19, but risk stratification of confirmed COVID-19 subjects remains unexplored. As such, we propose a model, DeepCOVID-Fuse, which fuses clinical variables from electronic health record (EHR) data and image features from chest X-rays to categorize infected COVID-19 patients into low, intermediate, and high-risk classes. Previous work on fusion and risk prediction includes utilizing the fusion of different image feature representations for COVID detection (14), tackling binary risk prediction with patient characteristic data (15), and predicting the chance of survival and kidney injury with tabular clinical and biochemistry data (16). Our DeepCOVID-Fuse model ensembles three different architectures with image and tabular clinical data, providing accurate fine-grained risk predictions on COVID-19 patients. Notably, we comprehensively compare the performance of fusion models trained on multiple modalities but tested on one or subset of modalities and find that testing the fusion model even with a missing modality still provides more informative predictions than networks trained on single modality.

## Materials and Methods

**Patients:**

This study, which included 2085 patients from more than 20 different sites across Northwestern Memorial Health Care System, was approved by the Northwestern institutional review board (STU00212323). All patients were tested positive from February 2020 to April 2020 and their corresponding electronic health records (EHR) were collected with a positive reverse transcription polymerase chain reaction (RT-PCR). Only the initial CXR after the first inpatient admission of each COVID-19 positive subject was selected in this study. Each of the COVID-positive patients was categorized into low, intermediate, or high risk classes. These three classes correspond respectively to 1) hospital length of stay (LOS) of less than one day, 2) hospital LOS greater than one day but no death or admittance to the ICU, and 3) death or admittance to the ICU according to their EHRs.

**CXRs Acquisition and Preprocessing:**
CXRs images were acquired from Northwestern Memorial Health Care System from over 20 sites and were preprocessed according to metadata with appropriate windowing operations. The grayscale images were first converted to 3-channel RGB images (with identical R, G, and B planes) as is the typical input of deep learning models. To remove unnecessary background and focus more on lung features, images were then center cropped using a UNet-based algorithm (17), which was pre-trained from the public CXRs dataset (18,19) to segment lung fields. Finally, all cropped images were resized to a resolution of 224×224 pixels, scaled to a range of 0 to 1 by dividing by 255, and normalized using ImageNet's mean and standard deviation before feeding them into the model. The preprocessing was applied on all training, validation, and test sets.

**Clinical Data Processing:**
Clinical variables were obtained from each subject's electronic health records (EHR) from different categories: basic demographic information, laboratory results, comorbidities, electrocardiogram (ECG), and modified early warning score (MEWS). First, each subject's first CXR was matched to its temporally closest EHR within 24 hours. We discarded features that were missing for more than 40% of the training cohort. Next, the features were classified into three types for preprocessing, namely, binary, categorical, and continuous, as shown in Supplementary Figure 1. Specifically, for binary features, such as comorbidities, if a value was missing from the training set, it was set to non-exist. For multi-class features, such as race and smoking status, missing values were set to an additional unknown class, and all classes were converted to one-hot vectors. For continuous features, the mean computed from the training set was used as the missing values and all features were scaled to the range of 0 to 1 using min-max normalization. The mean value for each clinical feature on the training set was applied to the validation and test sets. The details of all selected clinical features are provided in Supplementary Table 2.

**Model Details :**
DeepCOVID-Fuse is a weighted ensemble of three fusion neural network architectures trained for three-class classification (i.e., low risk, intermediate risk, and high risk) as shown in Figure 1, of which each individual network has two branches, including the CXR image branch and the clinical variable branch. In DeepCOVID-XR (4), six different networks were ensembled: DenseNet-121 (20), ResNet-50 (21), InceptionV3 (22), Inception-ResNetV2 (23), Xception (24), and EfficientNet-B2 (25). They all showed similar accuracy in binary COVID-19 classification, and ensembling multiple networks showed a diminishing return on accuracy versus efficiency. As such, for this work, three CNNs are chosen as the CXR image branch to process 224x224 chest X-rays, namely EfficientNet-B2, ResNet50, and DenseNet-121. For each network, a fully-connected layer was added to vary the feature dimension of the image branch, followed by a dropout layer.

Specifically, for the clinical variable branch, there was a corresponding fully-connected layer to process 99 clinical features from tabular electronic health record (EHR) data, followed by a dropout layer. The two branches were fused together using a concatenation layer, followed by two fully-connected layers and a three-class output node with a softmax activation function for final classification. Additional details about the hyperparameters are included in Supplementary materials. The training process of the entire framework for the two network branches consists of two steps. First, the weights of the image branch were initialized with the corresponding weights from DeepCOVID-XR while the clinical variable branch was randomly initialized as per TensorFlow standards. The convolutional layers of the image branch were frozen while only the clinical variable branch and the fusion layers were trained. Next, after early stopping ended the first stage of training, all layers were unfrozen and were fine-tuned in stage two. The outputs of each of the three models were averaged for the final prediction.

**Statistical Analysis:**

The overall accuracy, precision, recall, F1 score, and AUC (area under the receiver operating characteristic (ROC) curve) were calculated for comparison of different models, and 95% confidence intervals were obtained for each experiment among five independent experiment runs. McNemar's test (26) was performed for pairs to compare the accuracy, precision, recall, and F1 score, and DeLong test (27) was performed to compare the AUCs of different models. A p-value <0.05 was considered statistically significant. All analyses were conducted using packages Scikit-Learn and Scipy in Python 3.7.

# Results

**Experimental Design**

A total of 2085 subjects were included in this study. The demographic distribution details are shown in Supplementary Table 2. A three-class outcome was predicted for each COVID-19 subject, which are low risk (L), intermediate risk (I) and high risk (H). Since only the initial CXR after each COVID-19 subject's first inpatient admission was considered, all experiments had a total of 2085 images, of which 1657 (L:476, I:663, H:518; Mean age: 58.30 years±17.74 [standard deviation]; Female: 807) were used for training and 428 (L:119, I:176, H:133; 56.41 ± 17.03; 190) for validation, the same cohort applies to clinical features. A separate hold-out test set of 439 subjects (L:101, I:193, H:145; 56.51 ± 17.78; 205) from a different hospital were used to evaluate model performances.

Overall, three types of models were evaluated and compared in this study, including 1) the fusion models trained on CXRs and clinical features with different feature size combinations (*feat_dim* in Figure 1) from the image and feature branches, 2) the same models trained on CXRs only with the image branch and 3) the models trained on clinical features only with the feature branch. Additionally, for 1), we compared model performances of three fusion models individually and as an ensemble of them. For 2), to show the effect of the fusion model, we evaluated the fusion model on CXRs only. Likewise, for 3), we evaluated the fusion model on clinical features only, and compared the result with those trained directly on some machine learning algorithms. More experimental details are included in **Supplementary materials**. All experiments were run independently 5 times to account for model stochasticity. The models were trained and evaluated using one GPU (NVIDIA TITAN V) with Tensorflow 2.0 in Python 3.6.

**Performance of DeepCOVID-Fuse**

The performance of each individual fusion model and an ensemble of all models on the testing set are compared in Table 1. Overall, the ensemble model significantly outperformed all individual models on this COVID-19 risk prediction task, achieving an accuracy of 0.658, a recall of 0.660, a precision of 0.689, an F1 of 0.660 and an AUC score of 0.842. Notably, for all individual models, ablation studies of different feature dimensions from CXRs and clinical variables showed that models with higher proportional features from the clinical branch than the CXR branch (i.e., CXRs: clinical = 64:128) achieved better model predictions than equal (i.e., 128:128) or lower fractions (i.e., 1408:128). Furthermore, the fusion model with a DenseNet architecture had the best performance of AUC from 0.814 to 0.824, followed by a ResNet architecture from 0.794 to 0.815 and an EfficientNet architecture from 0.794 to 0.805.

**Comparison of Image-only with Fusion-image-only**

To show the importance of fusion, the performances of the model trained and tested on CXR image-only (Image-only) and the model pre-trained on the fusion model but tested on CXR image-only (Fusion-image-only) are provided in Table 2. Combined with the results in Table 1, the fusion model with additional clinical variables significantly improved COVID-19 risk prediction. The results further showed that for the same CXR-only test set, the pre-trained fusion model achieved a better model performance than the image-only model, even when clinical variables were not available. Specifically, for three individual models, the pre-trained fusion model improved the accuracy by 0.008~0.011, the recall by 0.004~0.012, the precision by 0.008~0.043, the F1 by 0.007~0.013 and the AUC by 0.009~0.016. Heatmaps generated from Image-only and Fusion-image-only models using gradient class activation maps (Grad-CAM) are provided in Figure 2 to visualize the salient features of each CXR from COVID-19 level classification. If models made the correct risk-level prediction, the heatmaps highlighted the abnormalities in the lungs. However, in some cases, when the Fusion-image-only model made the correct classification while the Image-only model misclassified, the heatmaps highlighted different feature patterns, where the former was in the lungs and the latter was not.

Typically, the clinical features are partially present. Supplementary Table 1 illustrates the results of the pretrained fusion model on CXR images, where the clinical variables are proportionally increased in the fusion model. The results showed that the more clinical features fused with the model, the better the performance of the pre-trained fusion model on the test set. For example, if 80% of the clinical variables are present, and only 20% are missing at random, the fusion model (e.g., with the DenseNet architecture) achieved an accuracy of 0.645, a recall of 0.647, a precision of 0.656, an F1 of 0.644 and an AUC of 0.816.

**Comparison of Feature-only with Fusion-Feature-only**

As shown in Table 3, to show the same effect of the fusion model on clinical features, we compared the performance of models trained and tested on clinical features only (Feature-only) with models first pre-trained on the fusion model but tested on the clinical features (Fusion-Feature-only). The result showed that even without CXRs as input, the pre-trained fusion model significantly outperformed the Feature-only model with an AUC of 0.733. In addition, we compared the neural network-based models with several machine learning algorithms trained on the same clinical features, including random forests (RM), quadratic discriminant analysis (QDA) and Linear Ridge classification (LR). Interestingly, RF achieved the best performance, followed by Fusion-Feature-only and LR, while Feature-only had the lowest performance among all metrics. Furthermore, the model in Table 1 combining CXRs and clinical variables still outperformed all results in Table 3.

## Discussion

In this study, we proposed a fusion model, DeepCOVID-Fuse, that incorporates clinical variables with CXRs to predict future risk of clinically meaningful outcomes in patients diagnosed with confirmed COVID-19. The fusion model was trained and tested using only the first inpatient admission data of each subject, which has great clinical implications for improving our healthcare management system, especially in intensive care units. The fusion model achieved an overall accuracy of 0.658 and an AUC of 0.842 on a hold-out testing set from a separate hospital. We further compared the fusion model with models trained on CXRs images only and clinical variables only, and evaluated the fusion model when only CXRs images or clinical variables were available respectively. To the best of our knowledge, we are the first to demonstrate the effectiveness of a fusion model, which was pre-trained on multiple modalities but is capable to have a better prediction performance and generate meaningful visual heatmaps when only one or parts of the modalities were available.

This is a three-class prediction (i.e., low, intermediate, and high risk) where the level is determined according to each subject's mortality status, mechanical ventilation, ICU admission and hospital length of stay (LoS). As the demand for hospital capacity is reported to be dramatically increasing during the COVID-19 pandemic (28), predicting ventilator usage or ICU admissions in advance will reduce pressure on hospitalization management. In addition, LoS is critical to the allocation of bed capacity so we chose a 1-day LoS as the separation to differentiate low and intermediate risk as only patients with a LoS of more than 1 day needed to be allocated a bed. Furthermore, the results in Table 1 and Table 2 show that the fusion model with the addition of clinical variables significantly improved risk performances over the model trained only on CXRs, indicating clinical variables are strongly associated with COVID-19 severity. Meanwhile, the performance of the ensemble fusion models outperforming each model individually is consistent with the previous study that the ensemble model reduces the generalization error of predictions (4).

In most real-world scenarios, the dataset is often missing or incomplete. As such, the fusion model is not guaranteed to feed in input from all modalities, i.e., some COVID-19 patients have either CXRs or a subset of clinical data. One study showed that this is a limitation of fusion models as predictions can be overly influenced by the most feature-rich modalities leading to poor generalization (29). However, our study shows that on input with only one or parts of modalities, using pre-trained fusion models can still achieve better performances than models trained on this one or part of modalities, as shown in Table 2 and Table 3. Learning correlations across different modalities is the possible explanation for this better performance. Specifically, since different modalities of a fusion model are simultaneously back-propagated through the loss, they complement each other, so the fusion model is able to learn better feature representations for each model branch. Therefore, even if only a subset of CXRs or clinical variables are available, fusion models still play an important role in learning more discriminative features. The experiments in Supplementary Table 1 further show that once the fusion model was pre-trained, the model performance continued to improve as long as more clinical variables were available in the test set. This has significant implications in future medical research, as it gives a strong support that if images are provided with more usable information during training, i.e., simple features such as age and gender, even if only images are available at testing stage, a better classification prediction can be achieved compared with that using only images to train and test models.

Heatmaps generated by Grad-CAM demonstrated from another perspective that fusion models can better learn feature representations of CXR images than image-only models. As shown in Figure 2e-2h, when only CXR images were available, the image-only model misclassified a high-risk subject as intermediate while the pre-trained fusion model made the correct prediction, which can be seen from their heatmaps, where the fusion one highlighted discriminative features of the lung but the other one located the wrong area. When making the correct predictions, as shown in Figure 2a-2d,2f-2h, all heatmaps looked at the close areas of the lung.

Although previous studies have existed to predict the severity of confirmed COVID-19 patients, our work has uniqueness and different focus in many ways. For example, Liang et al. developed a DL-based survival model on 1D clinical dataset collected at admission to predict the risk of COVID-19 patients being critically ill within 30 days, achieving an overall AUC above 0.85 (30). However, they were limited by lack of clinical dataset and no imaging data were available. Shamout et al. later proposed a deep learning model using CXRs and routine clinical variables to predict the deterioration risk (i.e., intubation, ICU admission or mortality) in COVID-19 subjects within 96 hours with an AUC of 0.786 (31). Similarly, Jiao et al. used a DL network combining CXR and clinical data to predict binary outcomes of COVID-19 patient severity (i.e., severe or not) and obtained AUCs ranging from 0.731 to 0.792 (32). Although two modalities were provided, both studies adopted a late fusion strategy with two independently trained models. By contrast, we trained an end-to-end fusion model that could learn and transfer information between two modalities. The only study similar to our work that combined initial CXRs and clinical variables into an end-to-end fusion model to predict mortality in COVID-19 subjects achieved an AUC of 0.82 (33). However, their model was only trained on 499 subjects aged in range 21 to 50, which may lead to poor model generalization, whereas we had 2085 subjects of any age. Most importantly, the focus of this paper is to comprehensively evaluate the statistical and visual performance of fusion models trained on multiple modalities but tested on one or a subset of modalities.

This study has some limitations. First, several clinical data are still missing or incomplete in the training dataset. Although we have shown that it is not necessary to have all clinical data in the test set, a more complete training dataset guarantees a better and more robust model. Second, we did not compare the performance of our fusion model with radiologists because risk prediction by experts on both CXR and clinical data is challenging and subjective. There is no true, universal ground truth.Finally, as shown in Table 3, we found that basic machine learning algorithms, such as random forests, achieved a better performance than deep learning-based models, indicating that our fusion model has not yet perfectly extracted features from 1D clinical data. Therefore, our future work would interface random forests with deep neural networks to further improve model performances.

In conclusion, we proposed a fusion model, DeepCOVID-Fuse, to predict risk levels in COVID-19 subjects using CXRs and clinical variables obtained at their initial inpatient admission. We showed that models combining both CXRs and clinical features outperformed models with only CXRs or features. Furthermore, we demonstrated that the pre-trained fusion model was able to achieve good model performance when only one or partial modality were available. We believe that this work, on one hand, demonstrates that it is possible to predict high-risk patients at admission to further benefit hospital triage systems. On the other hand, it has the potential to promote the use of fusion models in other fields of medical research.

## Supplemental Materials

**DeepCOVID-Fuse**
The architecture of the ensemble fusion model is shown in Figure 1. The ensemble was chosen because studies have shown that the ensemble architecture is able to improve model performance and generalization ability [4]. Different combinations of feature dimensions from the image branch and the clinical variable branch when concatenated were compared. The feature dimension of the clinical variable branch after a fully-connected layer was fixed at 128 while the feature dimension of the image branch varied, including 64 (smaller than clinical features), 128 (equal to clinical features) and the same dimension after each model's average pooling layer (larger than clinical features).

**Image-only model and Feature-only model**
The image-only model trained and tested only on CXRs images, which used the image branch of the fusion model. As shown in Figure 1, after the dropout layer of the image branch, the features were fed into the dense layer without the concatenation layer. Similarly, the feature-only model trained and tested only on clinical variables, which used the clinical branch of the fusion model, i.e., after the dropout layer of the clinical branch, the features were fed into the dense layer without the concatenation layer.

**Fusion-image-only and Fusion-feature-only model**
The Fusion-image-only model used the pre-trained DeepCOVID-Fuse model (i.e., the trained weights) but only used CXRs as input for testing. Although no clinical variables were used, the fusion model required them as input, so we computed the average clinical features from the training set and used them as input to the test set, which means all test sets had the same clinical features calculated from the training set.
The Fusion-feature-only model used the same pre-trained DeepCOVID-Fuse model but only used clinical features as input for testing. The fusion model still required a CXR input, so we chose a CXR image from the training set that had the most equal probability for the three classes predicted by the model. In other words, we didn't want the images to add any bias to the fusion model predicted from clinical features alone.

**Model Implement Details**
Each individual architecture first initialized the weights from [4] (GitHub: https://github.com/IVPLatNU/deepcovidxr). The convolutional layers were frozen, and the model was trained to finetune all fully-connected layers and the clinical variable branch. Then, all layers were unfrozen and trained again on the training set. We used stochastic gradient descent as the optimizer with initial learning of 0.0002 and momentum of 0.9. The batch size was set as 16 and the early stopping with a patience of 8 was used to avoid overfitting. The loss function was the class-weighted categorical cross-entropy loss, which was defined as:

$$L = \sum_i - \alpha_i y_i log(p_i), \tag{1}$$

where $y_i$ is the ground truth class of each subject $i$, $p_i$ is the predicted probabilities from the model and $\alpha_i$ is the class weight, which is calculated by inverting the frequency of each class to alleviate the class imbalance problem. Our code is provided freely for public on GitHub at #[the link for the codes].

# References

1. WHO Coronavirus (COVID-19) Dashboard. Available from: https://covid19.who.int

2. Emergency use authorization (EUA) summary COVID-19 RT-PCR test (laboratory corporation of America). 2022.

3. Sverzellati N, Ryerson CJ, Milanese G, Renzoni EA, Volpi A, Spagnolo P, et al. Chest radiography or computed tomography for COVID-19 pneumonia? Comparative study in a simulated triage setting. Eur Respir J. 2021 Sep 1 [cited 2022 Aug 4];58(3).

4. Wehbe RM, Sheng J, Dutta S, Chai S, Dravid A, Barutcu S, et al. DeepCOVID-XR: An Artificial Intelligence Algorithm to Detect COVID-19 on Chest Radiographs Trained and Tested on a Large U.S. Clinical Data Set. Radiology. 2021 Apr;299(1):E167–76.

5. Oh Y, Park S, Ye JC. Deep Learning COVID-19 Features on CXR Using Limited Training Data Sets. IEEE Trans Med Imaging. 2020 Aug;39(8):2688–700.

6. Harmon SA, Sanford TH, Xu S, Turkbey EB, Roth H, Xu Z, et al. Artificial intelligence for the detection of COVID-19 pneumonia on chest CT using multinational datasets. Nat Commun. 2020 Aug 14;11(1):4080.

7. Wynants L, Calster BV, Collins GS, Riley RD, Heinze G, Schuit E, et al. Prediction models for diagnosis and prognosis of covid-19: systematic review and critical appraisal. BMJ. 2020 Apr 7;369:m1328.

8. ACR Recommendations for the use of Chest Radiography and Computed Tomography (CT) for Suspected COVID-19 Infection [Internet]. [cited 2022 Aug 4]. Available from: https://www.acr.org/Advocacy-and-Economics/ACR-Position-Statements/Recommendations-for-Chest-Radiography-and-CT-for-Suspected-COVID19-Infection

9. Castiglioni I, Ippolito D, Interlenghi M, Monti CB, Salvatore C, Schiaffino S, et al. Machine learning applied on chest x-ray can aid in the diagnosis of COVID-19: a first experience from Lombardy, Italy. Eur Radiol Exp. 2021 Feb 2;5:7.

10. Ozturk T, Talo M, Yildirim EA, Baloglu UB, Yildirim O, Rajendra Acharya U. Automated detection of COVID-19 cases using deep neural networks with X-ray images. Comput Biol Med. 2020 Jun 1;121:103792.

11. Wang L, Lin ZQ, Wong A. COVID-Net: a tailored deep convolutional neural network design for detection of COVID-19 cases from chest X-ray images. Sci Rep. 2020 Nov 11;10(1):19549.

12. Hemdan EED, Shouman MA, Karar ME. COVIDX-Net: A Framework of Deep Learning Classifiers to Diagnose COVID-19 in X-Ray Images [Internet]. arXiv; 2020 [cited 2022 Aug 11]. Available from: http://arxiv.org/abs/2003.11055

13. DeGrave AJ, Janizek JD, Lee SI. AI for radiographic COVID-19 detection selects shortcuts over signal. Nat Mach Intell. 2021 Jul;3(7):610–9.

14. Bayram F, Eleyan A. COVID-19 detection on chest radiographs using feature fusion based deep learning. Signal Image Video Process. 2022 Sep 1;16(6):1455–62.

15. Quiroz-Juárez MA, Torres-Gómez A, Hoyo-Ulloa I, León-Montiel R de J, U'Ren AB. Identification of high-risk COVID-19 patients using machine learning. PLOS ONE. 2021 Sep 20;16(9):e0257234.

16. Aboutalebi H, Pavlova M, Shafiee MJ, Florea A, Hryniowski A, Wong A. COVID-Net Biochem: An Explainability-driven Framework to Building Machine Learning Models for Predicting Survival and Kidney Injury of COVID-19 Patients from Clinical and Biochemistry Data [Internet]. arXiv; 2022 [cited 2022 Aug 4]. Available from: http://arxiv.org/abs/2204.11210

17. Ronneberger O, Fischer P, Brox T. U-Net: Convolutional Networks for Biomedical Image Segmentation [Internet]. arXiv; 2015 [cited 2022 Aug 4]. Available from: http://arxiv.org/abs/1505.04597

18. Jaeger S, Candemir S, Antani S, Wáng YXJ, Lu PX, Thoma G. Two public chest X-ray datasets for computer-aided screening of pulmonary diseases. Quant Imaging Med Surg. 2014 Dec;4(6):47577–47477.

19. Shiraishi J, Katsuragawa S, Ikezoe J, Matsumoto T, Kobayashi T, Komatsu K ichi, et al. Development of a Digital Image Database for Chest Radiographs With and Without a Lung Nodule. Am J Roentgenol. 2000 Jan;174(1):71–4.

20. Huang G, Liu Z, van der Maaten L, Weinberger KQ. Densely Connected Convolutional Networks [Internet]. arXiv; 2018 [cited 2022 Aug 11]. Available from: http://arxiv.org/abs/1608.06993

21. He K, Zhang X, Ren S, Sun J. Deep Residual Learning for Image Recognition [Internet]. arXiv; 2015 [cited 2022 Aug 11]. Available from: http://arxiv.org/abs/1512.03385

22. Szegedy C, Vanhoucke V, Ioffe S, Shlens J, Wojna Z. Rethinking the Inception Architecture for Computer Vision. In: 2016 IEEE Conference on Computer Vision and Pattern Recognition (CVPR) [Internet]. Las Vegas, NV, USA: IEEE; 2016 [cited 2022 Aug 11]. p. 2818–26. Available from: http://ieeexplore.ieee.org/document/7780677/

23. Szegedy C, Ioffe S, Vanhoucke V, Alemi A. Inception-v4, Inception-ResNet and the Impact of Residual Connections on Learning [Internet]. arXiv; 2016 [cited 2022 Aug 11]. Available from: http://arxiv.org/abs/1602.07261

24. Chollet F. Xception: Deep Learning with Depthwise Separable Convolutions. In: 2017 IEEE Conference on Computer Vision and Pattern Recognition (CVPR) [Internet]. Honolulu, HI: IEEE; 2017 [cited 2022 Aug 11]. p. 1800–7. Available from: http://ieeexplore.ieee.org/document/8099678/

25. Tan M, Le QV. EfficientNet: Rethinking Model Scaling for Convolutional Neural Networks [Internet]. arXiv; 2020 [cited 2022 Aug 11]. Available from: http://arxiv.org/abs/1905.11946

26. McNemar Q. Note on the sampling error of the difference between correlated proportions or percentages. Psychometrika. 1947 Jun;12(2):153–7.

27. DeLong ER, DeLong DM, Clarke-Pearson DL. Comparing the Areas under Two or More Correlated Receiver Operating Characteristic Curves: A Nonparametric Approach. Biometrics. 1988;44(3):837–45.

28. Sacchetto D, Raviolo M, Beltrando C, Tommasoni N. COVID-19 Surge Capacity Solutions: Our Experience of Converting a Concert Hall into a Temporary Hospital for Mild and Moderate COVID-19 Patients. Disaster Med Public Health Prep. 2022 Jun;16(3):1273–6.

29. Huang SC, Pareek A, Seyyedi S, Banerjee I, Lungren MP. Fusion of medical imaging and electronic health records using deep learning: a systematic review and implementation guidelines. Npj Digit Med. 2020 Oct 16;3(1):1–9.

30. Liang W, Yao J, Chen A, Lv Q, Zanin M, Liu J, et al. Early triage of critically ill COVID-19 patients using deep learning. Nat Commun. 2020 Jul 15;11(1):3543.

31. Shamout FE, Shen Y, Wu N, Kaku A, Park J, Makino T, et al. An artificial intelligence system for predicting the deterioration of COVID-19 patients in the emergency department. Npj Digit Med. 2021 May 12;4(1):1–11.

32. Jiao Z, Choi JW, Halsey K, Tran TML, Hsieh B, Wang D, et al. Prognostication of patients with COVID-19 using artificial intelligence based on chest x-rays and clinical data: a retrospective study. Lancet Digit Health. 2021 May 1;3(5):e286–94.

33. Kwon YJ (Fred), Toussie D, Finkelstein M, Cedillo MA, Maron SZ, Manna S, et al. Combining Initial Radiographs and Clinical Variables Improves Deep Learning Prognostication in Patients with COVID-19 from the Emergency Department. Radiol Artif Intell. 2021 Mar;3(2):e200098.

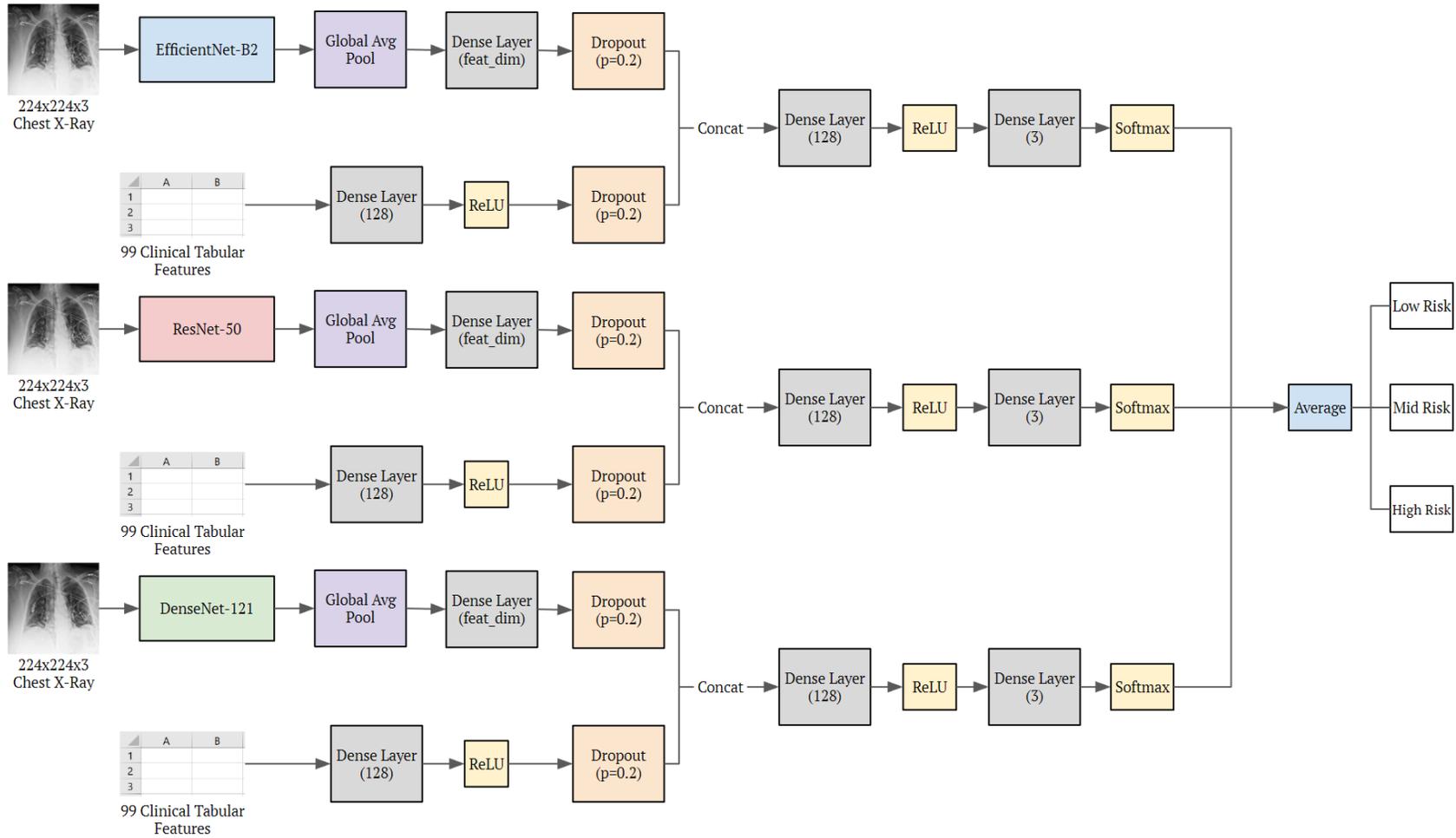

Figure 1. The architecture of DeepCOVID-Fuse ensemble model. The preprocessed image was fed into three different CNN architectures, followed by a fully-connected layer to transform the image feature dimension. The clinical tabular features were fed into a fully-connected layer. The features were fused and fed into another fully connected layer, followed by the last classification layer with softmax as the activation function. *Feat_dim* changes to compare different combinations of features from image and clinical feature branches.

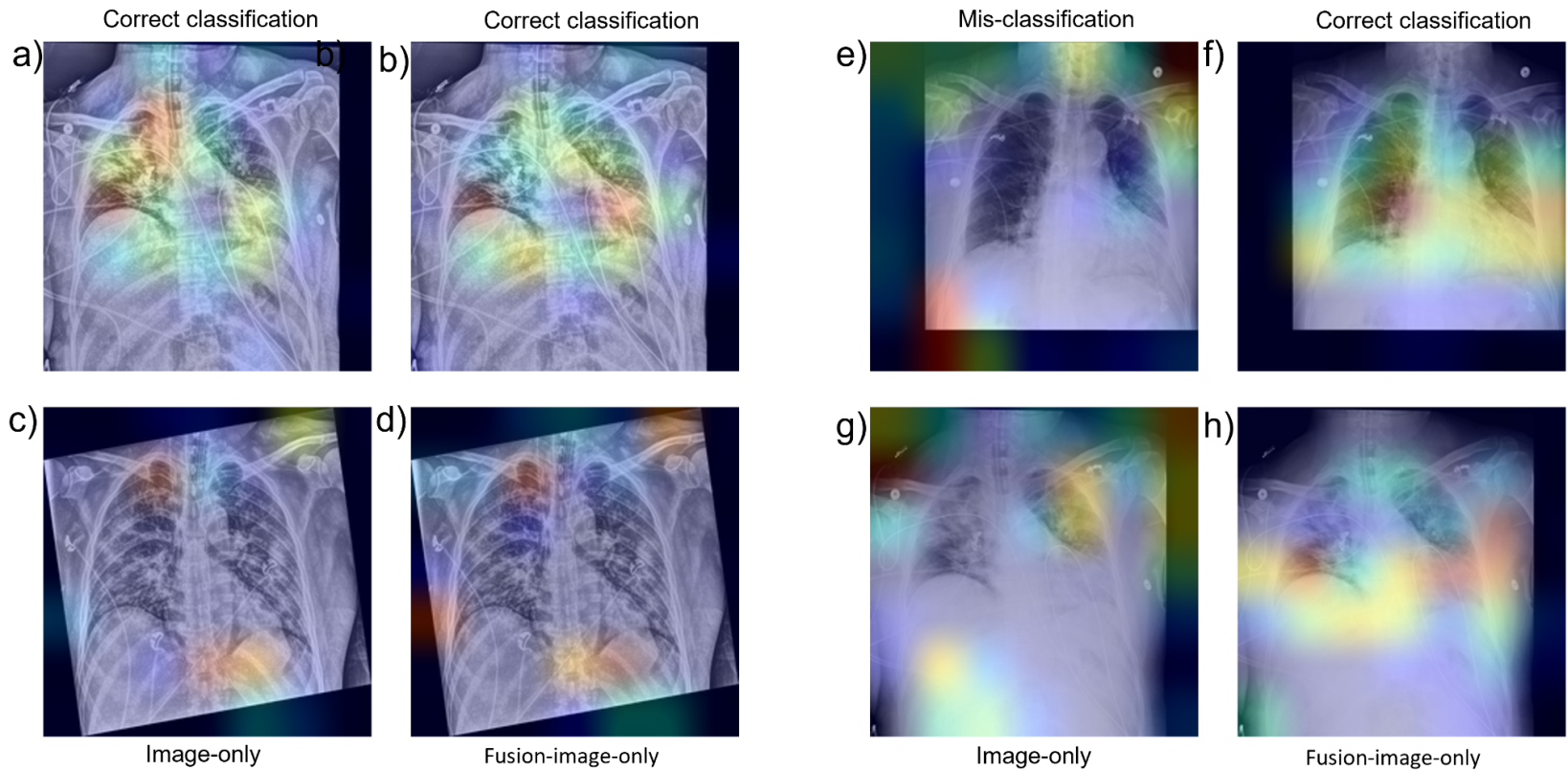

Figure 2: Heatmaps of Gradient class activation maps (Grad-CAM) generated from the model showing the location of important features for high risk predictions in confirmed COVID-19 patients. The redder the intensity of the heatmap, the more important the feature areas. Heatmaps generated from Image-only models (a,c,e,g) and Fusion-image-only models (b,d,f,h) are compared for cases of correct (a-d, f, h) and incorrect predictions (e, g). For the same subject using only CXR as model input, the Fusion model made correct predictions, with heatmaps (f, h) highlighting abnormalities in the lungs, while the image-only model misclassified the predictions, (e, g) highlighting unnecessary background.

Table 1. Performance of fusion models for risk predictions in confirmed COVID-19 subjects on external test sets with different combinations of latent feature sizes from X-rays and clinical variables.

| | EfficientNet | | | ResNet | | | DenseNet | | | Ensemble |
|---|---|---|---|---|---|---|---|---|---|---|
| Latent feature (x_ray*clinical data) | 64*128 | 128*128 | 1408*128 | 64*128 | 128*128 | 2048*128 | 64*128 | 128*128 | 1408*128 | 64*128 |
| Accuracy | 0.618 [0.600-0.637] | 0.622 [0.606-0.638] | 0.626 [0.590-0.662] | 0.628 [0.610-0.645] | 0.630 [0.620-0.642] | 0.611 [0.589-0.632] | 0.658 [0.650-0.667] | 0.638 [0.622,0.654] | 0.640 [0.632,0.647] | **0.658*** |
| Recall | 0.619 [0.600-0.639] | 0.622 [0.606-0.638] | 0.626 [0.590-0.662] | 0.626 [0.595-0.656] | 0.633 [0.623-0.642] | 0.611 [0.589-0.632] | 0.657 [0.649-0.666] | 0.638 [0.621-0.655] | 0.640 [0.632,0.647] | **0.660*** |
| Precision | 0.649 [0.631-0.666] | 0.648 [0.620-0.676] | 0.675 [0.648-0.702] | 0.665 [0.652-0.678] | 0.675 [0.664-0.685] | 0.652 [0.619-0.685] | 0.671 [0.658,0.684] | 0.641 [0.623-0.659] | 0.647 [0.635,0.659] | **0.689*** |
| F1 | 0.616 [0.599-0.633] | 0.619 [0.603-0.637] | 0.623 [0.583,0.663] | 0.626 [0.608-0.645] | 0.627 [0.612-0.642] | 0.607 [0.586-0.627], | 0.658 [0.650,0.666] | 0.638 [0.621-0.655] | 0.639 [0.632,0.647] | **0.660*** |
| AUC | 0.805 [0.798-0.812] | 0.794 [0.778-0.811] | 0.804 [0.780-0.827] | 0.815 [0.804-0.826] | 0.815 [0.809-0.820] | 0.794 [0.782-0.807] | 0.824 [0.822-0.826] | 0.814 [0.797,0.831] | 0.820 [0.805,0.836] | **0.842*** |

Note. - Data in parentheses are 95% CIs from five repeated experiment runs. AUC = area under the receiver operating characteristic curve, Latent feature = Image and clinical feature dimensions when concatenated in a fusion model
* P value < .05 denotes the comparisons are statistically significant.

Table 2. Performance of models for risk predictions in confirmed COVID-19 subjects on external test sets using only CXRs as model input.

| Covid-Level | EfficientNet image-only | ResNet image-only | DenseNet image-only | Image-only ensemble | EfficientNet fusion-image-only | ResNet fusion-image-only | DenseNet fusion-image-only | Fusion-image-only ensemble |
|---|---|---|---|---|---|---|---|---|
| Accuracy | 0.582 [0.572-0.591] | 0.614 [0.604,0.624] | 0.615 [0.608-0.622] | **0.621*** | 0.593 [0.581-0.606] | 0.625 [0.615-0.634] | 0.623 [0.604, 0.641] | **0.632*** |
| Recall | 0.581 [0.572-0.591] | 0.616 [0.604,0.624] | 0.616 [0.607-0.624] | **0.619*** | 0.593 [0.582-0.606] | 0.625 [0.615-0.634] | 0.620 [0.608, 0.632] | **0.629*** |
| Precision | 0.604 [0.594,0.614] | 0.664 [0.645, 0.683] | 0.631 [0.627-0.634] | **0.665*** | 0.657 [0.646-0.667] | 0.662 [0.643-0.681] | 0.639 [0.623,0.647] | **0.664*** |
| F1 | 0.576 [0.567-0.586] | 0.609 [0.595, 0.624] | 0.614 [0.606-0.621] | **0.620*** | 0.583 [0.567-0.600] | 0.619 [0.607-0.631] | 0.627 [0.611,0.639] | **0.634*** |
| AUC | 0.769 [0.764-0.774] | 0.798 [0.788, 808] | 0.781 [0.72-0.792] | **0.807*** | 0.781 [0.768-0.796] | 0.807 [0.803-0.811] | 0.797 [0.784,0.806] | **0.813*** |

Note. - Data in parentheses are 95% CIs from five repeated experiment runs. AUC = area under the receiver operating characteristic curve, image-only = models trained and tested with CXRs only, fusion-image-only = pre-trained fusion models but tested with CXRs only.
* P value < .05 denotes the comparisons are statistically significant.

Table 3. Performance of models for risk predictions in confirmed COVID-19 subjects on external test sets using only clinical variables as model input.

| Covid-Level | DNN feature-only | Fusion feature-only | Random Forests | QDA | Linear Ridge |
|---|---|---|---|---|---|
| Accuracy | 0.440 [0.432,0.448] | 0.539 [0.525, 0.553] | **0.560 [0.553,0.567]** | 0.526 [0.519, 0.533] | 0.536 [0.527,0.546] |
| Recall | 0.441 [0.430,0.449] | 0.540 [0.526,0.555] | **0.563 [0.554,0.569]** | 0.528 [0.517, 0.539] | 0.533 [0.525,0.541] |
| Precision | 0.193 [0.183,0.214] | 0.567 [0.553,0.582] | **0.588 [0.517,0.671]** | 0.532 [0.526, 0.538] | 0.544 [0.532,0.556] |
| F1 | 0.269 [0.253,0.280] | 0.560 [0.542,0.577] | **0.573 [0.568,0.581]** | 0.479 [0.461,0.496] | 0.536 [0.527,0.545] |
| AUC | 0.502 [0.481,0.522] | 0.733 [0.730,0.737] | **0.768 [0.759, 0.777]** | 0.600 [0.587, 0.613] | 0.625 [0.613,0.636] |

Note. - Data in parentheses are 95% CIs from five repeated experiment runs. AUC = area under the receiver operating characteristic curve, QDA = Quadratic Discriminant Analysis, feature-only = models trained and tested with clinical variables only, fusion-feature-only = pre-trained fusion models but tested with clinical variables only.

Supplementary Figure

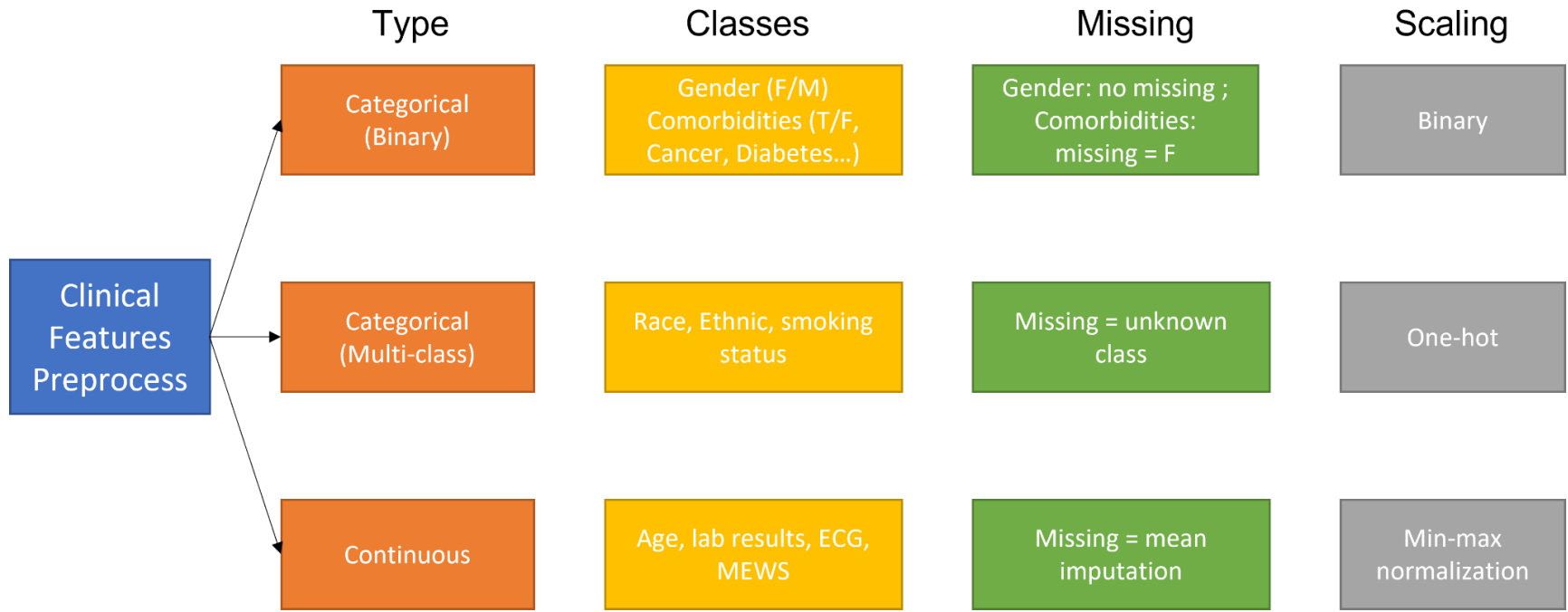

Figure 1: The preprocessing of clinical features. Features are classified into three types, binary, multi-class, and continuous with different missing imputation and scaling operations.

Supplementary Table 1. Performance of fusion-image-only models for risk predictions in confirmed COVID-19 subjects on external test sets using CXRs as model input with a random subset (%) of clinical variables (0: no clinical features, 100: full clinical features).

| Covid-Level (Proportion of clinical features in fusion) | 0 | 20% | 40% | 60% | 80% | 100% |
|---|---|---|---|---|---|---|
| Accuracy | 0.623 [0.604, 0.641] | 0.625 [0.615, 0.635] | 0.630 [0.620,0.640] | 0.635 [0.620, 0.650] | 0.645 [0.634, 0.656] | **0.658 [0.650-0.667]** |
| Recall | 0.620 [0.608, 0.632] | 0.623 [0.607,0.638]] | 0.628 [0.617,0.639]] | 0.634 [0.615,0.653] | 0.647 [0.635, 0.657] | **0.657 [0.649-0.666]** |
| Precision | 0.639 [0.623,0.647] | 0.637 [0.619,0.655]] | 0.640 [0.622, 0.658] | 0.647 [0.632, 0.662] | 0.656 [0.643, 0.669] | **0.671 [0.658,0.684]** |
| F1 | 0.627 [0.611,0.639] | 0.627 [0.612,0.642] | 0.630 [0.614, 0.645] | 0.634 [0.619, 0.649] | 0.644 [0.633,0.656] | **0.658 [0.650,0.666]** |
| AUC | 0.797 [0.784,0.806] | 0.798 [0.787,0.809] | 0.804 [0.796,0.812] | 0.808 [0.802, 0.814] | 0.816 [0.814, 0.818] | **0.824 [0.822,0.826]** |

Note. - Data in parentheses are 95% CIs from five repeated experiment runs. AUC = area under the receiver operating characteristic curve, fusion-image-only = pre-trained fusion models but tested with CXRs only.

Supplementary Table 2: Patient characteristics from the training, validation and test set.

| Clinical Variables | Training | Validation | Test |
|---|---|---|---|
| **Total (n)** | 1657 | 428 | 439 |
| **Gender (M/F)** | 850/807 | 238/190 | 234/205 |
| **Age (mean±std)** | 58.30±17.74 | 56.41 ± 17.03 | 56.51 ± 17.78 |
| **RACE(n)** | | | |
| Asia | 78 | 7 | 10 |
| Black or African American | 436 | 112 | 60 |
| White | 856 | 234 | 325 |
| Other | 234 | 64 | 32 |
| Unknown | 53 | 11 | 12 |
| **ETHNIC(n)** | | | |
| Hispanic or Latino | 381 | 127 | 256 |
| Not Hispanic or Latino | 1231 | 290 | 176 |
| Other | 45 | 11 | 7 |
| **SMOKING STATUS(n)** | | | |
| Never smoker | 877 | 245 | 300 |
| Smoker | 586 | 137 | 120 |
| Unknown | 199 | 46 | 19 |
| Atrial Rate | 91.27±31.95 | 95.45 ± 37.95 | 98.01 ± 35.62 |
| **ECG (mean±std)** | | | |
| P-R Interval | 164.05±26.45 | 155.22 ± 27.92 | 154.32 ± 27.90 |
| QRS Duration | 91.58±20.68 | 90.31 ± 19.11 | 88.48 ± 17.16 |
| QT | 376.35±49.57 | 369.93 ± 50.00 | 359.22 ± 41.55 |
| QTC | 448.81±35.93 | 444.39 ± 36.73 | 444.93 ± 30.49 |
| BP Diastolic | 76.89±14.85 | 75.75 ± 14.02 | 78.15 ± 14.12 |
| BP Systolic | 136.32±22.60 | 134.83 ± 22.63 | 135.87 ± 22.47 |
| Pulse | 93.76±18.93 | 96.30 ± 20.63 | 99.01 ± 18.45 |
| Respirations | 20.99±5.16 | 21.58 ± 5.24 | 23.40 ± 7.33 |
| SPO2 | 95.11±6.69 | 94.83 ± 5.09 | 92.45 ± 10.41 |
| Temperature | 99.15±1.50 | 99.32 ± 1.48 | 99.96 ± 1.61 |
| **COMORBIDITY(n)** | | | |
| Asthma | 4 | 2 | 0 |
| COPD | 249 | 38 | 24 |
| Cancer | 72 | 8 | 21 |
| Cardiovascular | 654 | 156 | 113 |
| Cerebrovascular disease | 285 | 65 | 49 |
| Chronic pulmonary disease | 543 | 131 | 96 |
| Congestive heart failure | 378 | 78 | 41 |
| Dementia | 177 | 43 | 55 |
| Diabetes | 614 | 170 | 191 |
| HIV_AIDS | 39 | 27 | 4 |
| Hemiplegia or paraplegia | 58 | 18 | 10 |
| Hypertension | 1026 | 251 | 240 |
| Immunological | 105 | 37 | 18 |
| Malignancy | 239 | 51 | 62 |
| Metastatic Solid Tumor | 126 | 25 | 26 |
| Myocardial infarction | 131 | 23 | 17 |
| Peptic ulcer disease | 90 | 15 | 16 |
| Peripheral vascular disease | 228 | 44 | 40 |
| Renal disease | 538 | 120 | 137 |
| Rheumatic disease | 74 | 31 | 13 |
| liver disease | 206 | 36 | 78 |
| **LAB (mean±std)** | | | |
| A EOS PERCENT | 0.84±1.66 | 0.63 ± 1.33 | 0.68 ± 1.52 |
| ABSOLUTE BASOPHILS | 0.01±0.03 | 0.01 ± 0.03 | 0.01 ± 0.03 |
| ABSOLUTE EOSINOPHILS | 0.05±0.12 | 0.05 ± 0.12 | 0.07 ± 0.43 |
| ABSOLUTE IMMATURE GRANULOCYTES, AUTOMATED | 0.05±0.10 | 0.04 ± 0.07 | 0.05 ± 0.09 |
| ABSOLUTE LYMPHOCYTES | 1.38±2.53 | 1.35 ± 0.90 | 1.22 ± 0.79 |
| ABSOLUTE MONOCYTES | 0.58±0.83 | 0.52 ± 0.26 | 0.56 ± 0.33 |
| ABSOLUTE NEUTROPHILS | 5.41±4.79 | 5.36 ± 3.34 | 5.98 ± 3.79 |
| ALBUMIN | 3.82±0.52 | 3.77 ± 0.47 | 3.72 ± 0.42 |
| ALKALINE PHOS | 79.84±49.65 | 74.65 ± 41.12 | 85.26 ± 55.43 |
| ALT (SGPT) | 39.80 59.28 | 43.36 ± 50.74 | 38.44 ± 35.16 |
| ANION GAP | 12.06±3.47 | 11.86 ± 3.48 | 10.72 ± 3.19 |
| AST (SGOT) | 48.17±89.00 | 48.68 ± 47.42 | 43.26 ± 38.67 |
| BASOPHILS | 0.19±0.41 | 0.18 ± 0.40 | 0.14 ± 0.35 |
| C-REACTIVE PROTEIN | 38.77±60.38 | 29.75 ± 47.82 | 31.93 ± 51.42 |
| CALCIUM | 8.91±0.60 | 8.93 ± 0.75 | 8.90 ± 0.59 |
| CHLORIDE | 100.68±5.04 | 101.07 ± 5.18 | 99.43 ± 5.38 |
| CO2 | 23.80±3.68 | 23.72 ± 3.65 | 24.83 ± 3.56 |
| CREATININE | 1.38±1.68 | 1.43 ± 1.42 | 1.23 ± 1.37 |
| D-DIMER | 1105.15±3456.19 | 1199.95 ± 5898.18 | 1635.64 ± 6804.57 |
| DIRECT BILIRUBIN | 0.21±0.20 | 0.20 ± 0.28 | 0.20 ± 0.21 |
| EOSINOPHILS | 0.82±1.57 | 0.77 ± 1.33 | 0.94 ± 1.66 |
| GFR(AFRICAN AMERICAN) | 70.13±35.81 | 66.05 ± 30.42 | 55.31 ± 12.02 |
| GFR(OTHERS) | 62.80±28.63 | 58.85 ± 24.83 | 53.65 ± 13.33 |
| GLUCOSE | 143.93±78.06 | 141.58 ± 70.19 | 155.08 ± 82.29 |
| HEMATOCRIT | 39.71±5.58 | 39.53 ± 5.52 | 39.69 ± 5.57 |
| HEMOGLOBIN | 13.05±2.04 | 13.00 ± 2.07 | 13.30 ± 2.08 |
| IMMATURE GRANULOCYTES | 0.62±0.99 | 0.50 ± 0.68 | 0.63 ± 0.64 |
| INR | 1.26±0.67 | 1.66 ± 1.84 | 1.25 ± 0.44 |
| LDH | 383.11±418.68 | 343.41 ± 146.85 | 332.39 ± 168.69 |
| LYMPHOCYTES | 19.79±11.10 | 20.34 ± 11.81 | 17.78 ± 10.07 |
| MCH | 28.98±2.58 | 28.85 ± 2.37 | 29.39 ± 2.45 |
| MCHC | 32.80±1.46 | 32.82 ± 1.54 | 33.44 ± 1.41 |
| MCV | 88.31±6.41 | 87.93 ± 5.86 | 87.91 ± 6.49 |
| MONOCYTES | 8.32±4.21 | 7.70 ± 3.75 | 7.70 ± 3.81 |
| MPV | 10.38±0.96 | 10.34 ± 0.90 | 10.32 ± 0.94 |
| NEUTROPHILS | 70.03±13.62 | 70.46 ± 13.30 | 72.91 ± 12.32 |
| PLATELET COUNT | 224.41±87.40 | 233.81 ± 94.25 | 227.47 ± 95.41 |
| POTASSIUM | 4.03±0.61 | 3.99 ± 0.56 | 3.81 ± 0.50 |
| PROCALCITONIN | 0.73±4.05 | 0.83 ± 5.65 | 0.81 ± 4.48 |
| PROTHROMBIN TIME (PT) | 14.53±8.17 | 19.39 ± 21.93 | 14.09 ± 5.17 |
| RDW | 13.88±1.99 | 13.84 ± 1.97 | 13.65 ± 1.80 |
| RED CELL COUNT | 4.53±0.74 | 4.52 ± 0.71 | 4.54 ± 0.72 |
| SODIUM | 136.54±4.20 | 136.66 ± 4.18 | 134.68 ± 4.53 |
| TOTAL BILIRUBIN | 0.61±0.44 | 0.55 ± 0.29 | 0.65 ± 0.63 |
| TOTAL PROTEIN | 7.12±0.76 | 7.06 ± 0.71 | 7.29 ± 0.66 |
| TROPONIN-I | 0.15±1.04 | 0.03 ± 0.05 | 0.04 ± 0.12 |
| UREA NITROGEN | 19.44±15.77 | 20.99 ± 15.93 | 19.98 ± 20.35 |
| WHITE BLOOD CELL COUNT | 7.73±11.04 | 7.33 ± 3.58 | 7.96 ± 5.15 |